\documentclass[aps,prl,twocolumn,amsmath,amsfonts,amssymb, braket, groupedaddress,floatfix]{revtex4-2}

\usepackage{graphicx}% Include figure files
\usepackage{dcolumn}% Align table columns on decimal point
\usepackage{bm}% bold math
\usepackage{hyperref}% add hypertext capabilities
\usepackage[usenames,dvipsnames]{color}
\usepackage[normalem]{ulem}
\usepackage{amsmath}
\usepackage{amssymb}
\usepackage{braket}

\begin{document}

%\preprint{APS/123-QED}

\title{Magnetic Behavior of Ferro-, Antiferro-, and Ferrimagnetic Systems in the Griffiths Phase: A Theoretical Study}

\author{Sumanta Mukherjee$^{1,*}$}

\affiliation{$^1$ Solid State and Structural Chemistry Unit, Indian Institute of Science, Bengaluru, Karnataka 560012, India\\}

\email{sm31081985@gmail.com}

\date{\today}

\begin{abstract}
In this report, we provide a theoretical framework for the magnetic behavior of the Griffiths phase, which, along with three-dimensional spin-1/2 Ising ferromagnetic systems, can be extended to antiferromagnetic as well as ferrimagnetic systems. We find that the magnetic behavior in the Griffiths phase of three-dimensional antiferromagnetic and ferrimagnetic systems is more unusual than that of conventional ferromagnetic systems. However, this study offers a possible framework for the identification of Griffiths phase behavior in three-dimensional antiferromagnetic and ferrimagnetic systems.
\end{abstract}

\maketitle

\section{Introduction}
The concept of the Griffiths phase (GP) was first introduced in a seminal paper \cite{ref-1} by R. B. Griffiths, where it was suggested that the magnetization ($M$) fails to be an analytic function of the applied field ($H$) at $H = 0$ over a wide temperature ($T$) range above the global ferromagnetic (FM) transition temperature ($T_c$) in a randomly diluted Ising ferromagnet \cite{ref-2,ref-3,ref-4,ref-5}. Subsequently, the phenomenon was theoretically understood \cite{ref-6,ref-7} by incorporating quenched disorder and randomness in $T_c$ within the framework of the Ginzburg–Landau–Wilson (GLW) theory of phase transitions \cite{ref-8}. The presence of random quenched disorder in a ferromagnetic matrix may lead to the formation of regions of finite volume with a certain probability of occurrence, where the effect of disorder is minimal; these can therefore be considered disorder-free (clean) regions, often referred to as rare regions \cite{ref-6,ref-9}. The rare region effect and its response to measurable quantities, such as magnetization, are suggested to be the origin of non-analyticity \cite{ref-6,ref-7,ref-10,ref-11}, as introduced in the original paper\cite{ref-1}. Since then, a large number of ferromagnetic samples have been found to exhibit Griffiths phase–like behavior in the temperature range ($T_c < T < T_g$) above the global FM transition temperature $T_c$, where $T_g$ is the temperature above which the system recovers its paramagnetic behavior \cite{ref-12,ref-13,ref-14,ref-15,ref-16,ref-17}. In most of these samples, the inverse susceptibility $\chi^{-1} = H/M$ exhibits a downward deviation\cite{ref-12,ref-13,ref-14,ref-15,ref-16,ref-17} from the usual linear Curie–Weiss (CW) paramagnetic behavior below $T_g$. The variation of $\chi^{-1}$ in the GP region ($T_c < T < T_g$) is generally described using a power-law form: $\chi^{-1} = (T - T_c^R)^{1- \lambda_{g}(T)}$\cite{ref-14,ref-15,ref-16,ref-17}. While $\lambda_{g}(T)$ is a temperature-dependent function, its value near the FM transition temperature is used to characterize the strength of the Griffiths phase and to estimate the deviation from CW behavior. While this behavior quantifies the extent of deviation from CW behavior and has been used in various contexts to identify the GP, it does not necessarily indicate the presence of a genuine Griffiths singularity\cite{ref-18}. Along with the deviation from linear paramagnetic behavior, the nature of the GP is reflected in the power-law variation of magnetization with applied field, the non-analyticity of magnetization at $H = 0$, the divergence of differential susceptibility at $H = 0$, and other related features\cite{ref-1,ref-6,ref-7}, as will be discussed in the forthcoming section. Furthermore, most of the GP studies reported in the literature focus on Ising ferromagnets. The presence of GP behavior in antiferromagnetic (AFM) systems is extremely rare\cite{ref-19,ref-20} due to the unknown nature of magnetization behavior in the GP region of an AFM sample\cite{ref-18}. While sporadic attempts have been made to understand quantum GP behavior in low-dimensional AFM systems \cite{ref-21,ref-22,ref-23,ref-24,ref-25,ref-26,ref-27}, the three-dimensional (3D) counterpart of the FM Ising system is not well documented\cite{ref-18}.\\

Therefore, in this report, we present the typical behavior of a randomly diluted Ising ferromagnet within the framework of GLW theory, random $T_c$ disorder, finite-size scaling\cite{ref-28}, and optimal fluctuation theory\cite{ref-6}. To incorporate the present theory of quantum GP based on energy gap distribution\cite{ref-7,ref-21,ref-26}, we provide a simplified yet comprehensive model that can be extended to understand the effects of quenched disorder in randomly diluted three-dimensional Ising FM, AFM, and ferrimagnetic (FiM) systems. These results suggest that a mere deviation of susceptibility from Curie–Weiss behavior may not guarantee the formation of a GP region. The behavior of magnetization in the GP region of ferromagnets is unusual and can be even more complex in AFM and FiM samples. Nevertheless, we propose a possible and useful framework to understand and identify true GP signatures in different systems, including FM, AFM, and FiM systems.\\

\section{Methods, Results and Discussion}
\subsection {Ferromagnetic system}
First, we provide a general description of dilution effects and the occurrence of the GP in spin-1/2 Ising FM systems, where the theoretical understanding is relatively comprehensive\cite{ref-6,ref-7,ref-29}. \textbf{Figure 1a} illustrates the effect of dilution on the clean FM transition of a system\cite{ref-1,ref-6,ref-16}. In this context, dilution implies that some sites, which would normally be occupied by a spin-1/2 electron, are randomly left unoccupied-for example, by replacing them with nonmagnetic ions. The fraction and extent of such random replacement or disorder is represented by the fraction $p$ in \textbf{Figure 1a}. For a clean system, $p=0$, and the system exhibits a clean transition temperature $T_c^0$. As $p$ increases, the FM transition temperature $T_c$ of the bulk sample decreases, as indicated by the blue region in \textbf{Figure 1a}. However, as pointed out by R. B. Griffiths, for any $p$ value between 0 and 1 (or close to 1), the magnetization fails to be an analytic function of the magnetic field at $H=0$ at temperatures between $T_c^0$ and $T_c$\cite{ref-1}. This temperature range is commonly referred to as the GP region, as shaded in yellow in \textbf{Figure 1a}, where the Griffiths temperature $T_g$ corresponds to the temperature at which the non-analyticity first appears $T_g = T_c^0$\cite{ref-14,ref-16,ref-17}. As the system enters a disorder-induced GP region from a clean paramagnetic state, the inverse susceptibility deviates downward\cite{ref-12,ref-13,ref-14,ref-15,ref-16,ref-17} from linear CW behavior, as shown in \textbf{Figure 1b}. Such deviations from linear CW behavior, and their suppression upon application of a magnetic field, have often been used as an indication of GP formation in various systems\cite{ref-12,ref-13,ref-14,ref-15,ref-16,ref-17}. However, as will be discussed in the forthcoming section, this downward deviation alone may not be sufficient to confirm the formation of a true GP in different samples.\\
\begin{figure}[t]
\begin{center}
\includegraphics[width=1.0\columnwidth]{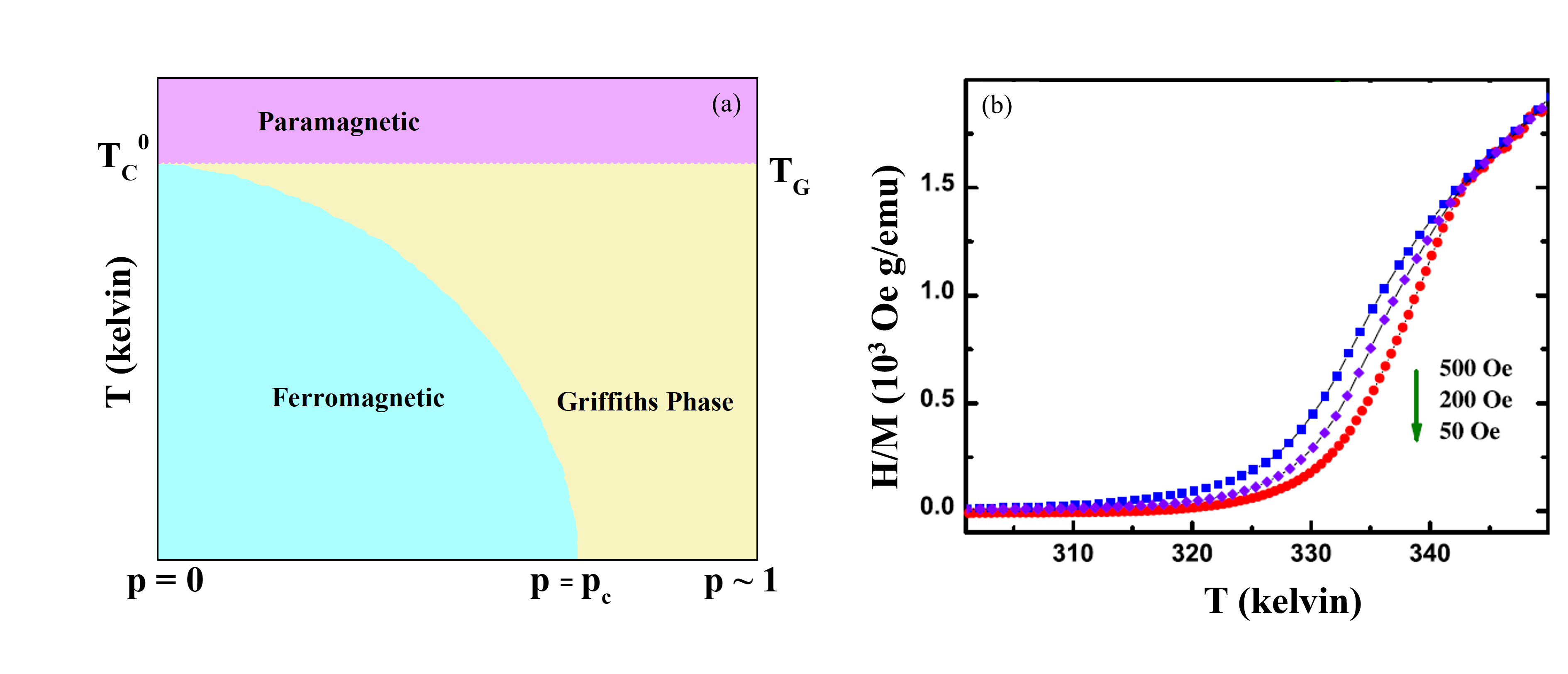}
\caption{(a) An approximate pictorial description of the random dilution effect on $T_c$ of a FM system\cite{ref-1,ref-6}. (b) Typical inverse suceptibility behavior of a FM sample in the GP region (reproduced with permission\cite{ref-16}).}
\label{fig1}
\end{center}
\end{figure}
\begin{figure}[t]
\begin{center}
\includegraphics[width=1.0\columnwidth]{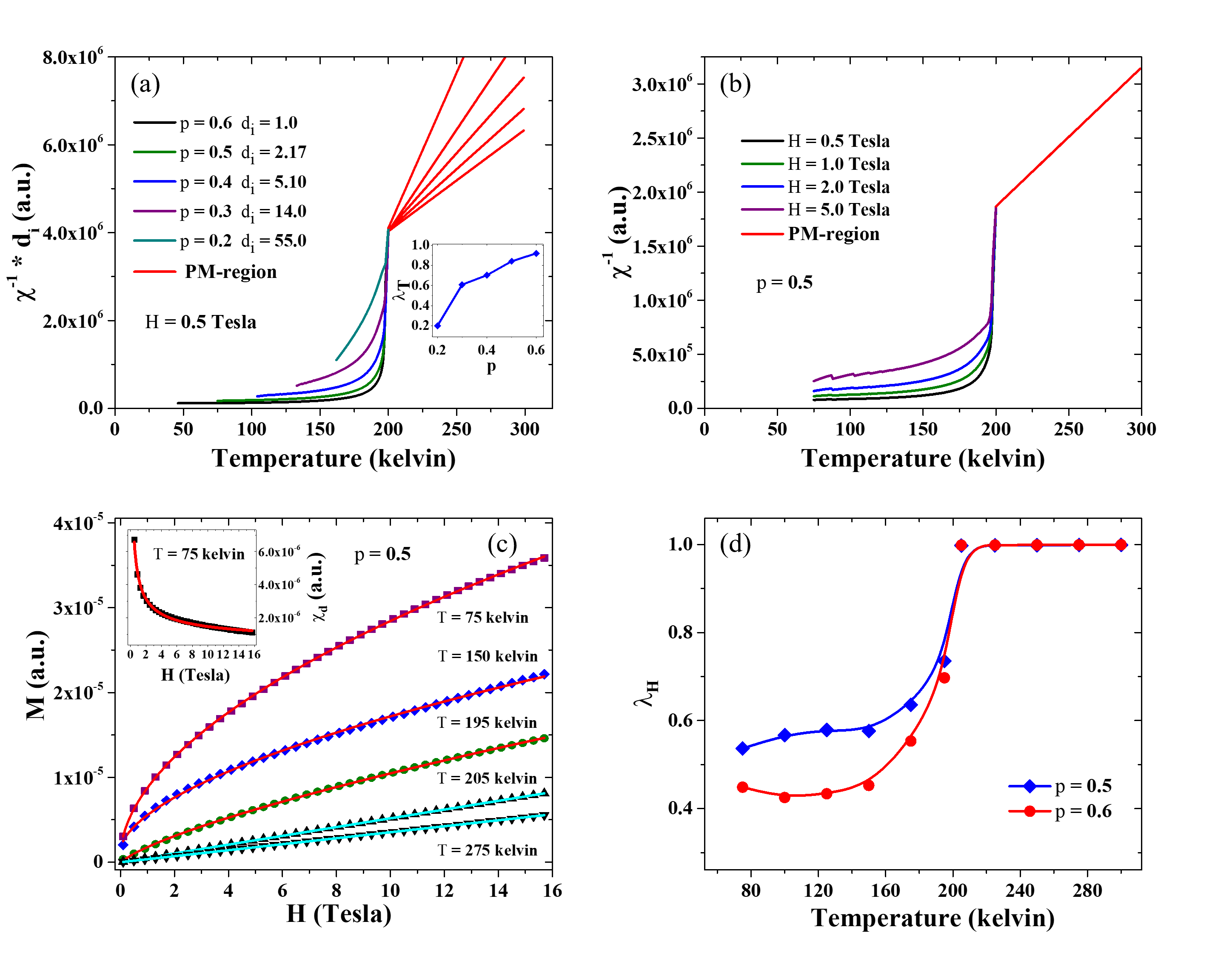}
\caption{(a) The calculated inverse susceptibility, $\chi^{-1}$, of a FM system, both above and below $T_g$, is plotted as a function of temperature for different disorder strengths, $p$. A magnetic field of $H = 0.5$ Tesla was used for these calculations. In order to compare the $\chi^{-1}$ values calculated at different disorder strength, $p$, the calculated $\chi^{-1}$ values were multiplied by a constant factor $d_i$ for better comparison. The inset of (a) shows the variation of $\lambda_g(T)$ as a function of disorder strength $p$; see the manuscript for details. Also note that, restricted by the maximum cluster size, the calculations were performed up to $\sim 198~\text{kelvin}$; the region between $\sim 198~\text{kelvin}$ and $200~\text{kelvin}$ is an extrapolation. (b) shows the variation of $\chi^{-1}$ at different fields for a fixed disorder strength of $p = 0.5$ (c) The calculated variation of $M$ with $H$ at a few selected temperatures for a fixed disorder strength of $p = 0.5$ is shown. The solid lines represent fits using a power-law behavior. (d) The estimated values of the exponent, obtained from the fitting of the power-law variation of $M$ versus $H$, are shown for two different disorder strengths.}
\label{fig2}
\end{center}
\end{figure}
To understand the effect of random dilutions, often introduced as quenched disorder in the GLW theory of phase transitions, one can consider a nonuniform and spatially dependent order parameter $\phi (x)$. In this case, the free energy $F$ within the GLW framework can be expressed as\cite{ref-6,ref-8}.
\begin{equation}
\resizebox{.9\hsize}{!}
{$F\{\phi(x)\} = \int d^3 x \left[ \frac{1}{2}\left\{\nabla \phi(x)\right\}^2 + \frac{r}{2}\left\{\phi(x)\right\}^2 + \frac{u}{4!}\left\{\phi(x)\right\}^4 - h \phi(x) \right]$}
\end{equation}
Where the quenched disorder is introduced by considering a distribution of $r$ as $r + \delta r(x)$. In other words, each point in space within this model has its own $T_{c}(x)$, or transition temperature, instead of a single global transition temperature\cite{ref-6,ref-8}. However, if the disorder is random, as in the case of random dilution, one may occasionally find regions in space of volume $V$ where the effect of disorder is minimal. These small regions, often referred to as rare regions, may then act as parts of a clean sample, each having its own transition temperature ($T_{c}^r$)\cite{ref-6,ref-9}. The probability that a rare region has a volume $V$ depends on the extent of disorder and is given by
$P_r = \exp {(-\beta V)}$, where $\beta = - \ln {(1-p)}$ \cite{ref-6}. In terms of the number of electrons $n$ within a rare region, the probability may be expressed as $P_r (n) \propto \exp {(-\beta n)}$\cite{ref-10,ref-11}. To express the $T_{c}^r$ of a particular rare region, one may use finite-size effects and random fluctuation theory, and define a distance $r_0 = (T - T_g)/T_g$. Within finite-size scaling, one may express the shift in $T_c$ of a rare region of length $L$ by defining $r^\prime (L)$ as\cite{ref-6,ref-28}.
\begin{equation}
r^\prime (L) = r_0 + \frac {K}{L^{1/\nu}}
\end{equation}
K is a constant, and for three-dimensional Ising spins $1/\nu = 1.56$\cite{ref-28}, the above expression can therefore be written in terms of the number of electrons as
\begin{equation}
r^\prime (n) = r_0 + \frac {K}{n^{0.52}}
\end{equation}
If $r^\prime (n)$ is negative for a rare region, the region will be ordered. In other words, for any $r_0$, the rare regions that are ordered will have
\begin{equation}
n \ge \left( \frac{K}{r_0}\right) ^{1/0.52}
\end{equation}
The equality sign determines the $T_{c}^r$ of a rare region. The value of $K$ is taken to be $1.0$ for all the calculations presented.\\
Within this method, one may now calculate the magnetization $M_t^g$ of the sample at any temperature and field below $T_g$ as
\begin{equation}
M_t^g = M_p^g + M_c^g
\end{equation}
Here, $M_p^g$ represents the contribution from the rare regions that are not ordered $(n = 1 \text{ to } m(T)-1)$ and can be calculated by solving the Weiss molecular field equation\cite{ref-30,ref-31} as
\begin{equation}
M_p^g = n_p^{gt} \mu_B \tanh \frac{\mu_B (H + \lambda_w M_p^g)}{k_B T}
\end{equation}
Here, the Weiss term $\lambda_w$ for the paramagnetic spins, which eventually order at a certain temperature, is determined by considering the global transition temperature $T_c$ as the ferromagnetic transition temperature.
For approximate calculations, and away from the percolation limit, one may crudely assume a linear suppression\cite{ref-32,ref-33} of $T_c$, given by $T_c = T_c^0 (1- p/p_c)$, where $p_c = 0.6884$ defines the percolation limit (3D, simple cubic lattice)\cite{ref-34} for the present calculations. On the other hand, the total number of paramagnetic electrons $n_p^{gt}$ is proportional to
\begin{equation}
n_p^{gt} \propto  \sum_{n=1}^{m(T)-1} n P_r (n)
\end{equation}
The contribution from the ordered rare regions can be expressed as
\begin{equation}
M_c^g (T,H) = \sum_{n=m(T)}^{n_{max}} P_r (n) M_c (n,H,T)
\end{equation}
Where,
\begin{equation}
M_c = n \mu_B \tanh \frac{\mu_B (H + \lambda_{wc} M_c)}{k_B T}
\end{equation}
Here, the Weiss term $\lambda_{wc}$ is determined by calculating the transition temperature of each cluster size ($T_c^r$) using the finite-size effect discussed previously. Note that, unlike the paramagnetic spins, the Weiss term here depends on the size of the clusters and therefore includes the finite-size effects\cite{ref-28} observed in clusters. Ideally, $n_{max}$ can be infinite; however, to make the calculations tractable, a maximum value of 7000 was used for $n_{max}$ in all cases.
Within this semi-classical framework for calculating cluster magnetization, the non-analyticity of the magnetization may not be fully captured. Therefore, we include a quantum description of cluster magnetization in our calculations\cite{ref-6,ref-7,ref-21,ref-29}, where each cluster is associated with an energy gap in its excitation spectrum, the magnitude of which decreases exponentially as the cluster volume increases\cite{ref-6,ref-7}. This behavior, together with the exponentially decreasing probability of the clusters, leads to a power-law distribution $(P(\epsilon))$ of energy gaps $\epsilon$ \cite{ref-6,ref-7,ref-21,ref-29}. This power-law distribution can be applied in various contexts to understand the nature of the GP in different samples. To utilize this distribution of gaps, and to make the calculation extendable to both AFM and FM systems, we have adopted a slightly different and more straightforward approach to incorporate the gap distribution into our magnetization calculations. For this, we write the exponentially increasing response of a cluster of size $n$ due to the exponential decrease of the excitation gap as $R_r(n) \propto \exp (\eta n)$. This, together with the probability function $P_r(n)$ , gives the overall response of a cluster of size $n$ to an observable as $P^\prime_r(n) = \exp [(\eta - \beta)n] \propto \exp (-\alpha \beta n)$. Furthermore, we consider that even for clusters of the same size, variations in their shape can introduce a distribution of effective length scales, and hence a distribution of gaps \cite{ref-9}. We further assume that such shape variations can provide a relevant source of gap fluctuations \cite{ref-9}. Although the dominant contribution to the gap distribution generally arises from the distribution of cluster sizes, shape fluctuations within the fixed-volume sector can still broaden the spectrum. Under suitable conditions, this resulting distribution of gaps may be approximated by an effective power-law form. Therefore, we adopt a power-law distribution and write the total magnetization from the ordered clusters in the following way, while changes in cluster size—and the corresponding shifts in gap values—are accounted for by adjusting the integration limits.
\begin{widetext}
\begin{equation}
M_c^g (T,H) = S(T)\sum_{n=m(T)}^{n_{max}} P^\prime_r(n) \int_{\epsilon_{min}(n)}^{\epsilon_{max}(n)} P_n(\epsilon) M_c (n,H,T) \tanh \frac{M_c (n,H,T) H}{\epsilon}
\end{equation}
\end{widetext}
Where,
\begin{equation}
\begin{aligned}
&M_c = n \mu_B \tanh \frac{\mu_B (H + \lambda_{wc} M_c)}{k_B T}\\
&P_n(\epsilon) = C(n) \epsilon^{(\frac{d}{z}-1)}\\
&\epsilon_{max}(n) = (300\epsilon_t)- q(\exp (-s/n))\\
&\epsilon_{min}(n) = (1\epsilon_t)- q(\exp (-s/n))
\end{aligned}
\end{equation}
Where the exponent $d/z$ was taken as $0.5$, the values of $q$ and $\epsilon_t$ are $0.003$, whereas a value of $350$ was used for $s$. Note that $\epsilon_{max}(n)$ may be approximately set by the behavior of $M_c (n,H,T) H/\epsilon$, whereas $\epsilon_{min}(n)$ can, in principle, be extended arbitrarily close to zero. However, to keep the numerical calculation tractable and to include the exponential decrease of $\epsilon$ with increasing cluster size, this specific dependence has been chosen. Furthermore, while these values provide the required power-law tail for the distribution of gaps and the exponential decrease of gaps with increasing cluster size, they can be modified depending on the type of calculation. $C(n)$ is used as a normalization constant for the probability distribution $P_n(\epsilon)$ and can be given, for $\frac{d}{z} = 0.5$ as:
\begin{equation}
C(n) = \frac {0.5}{\sqrt{\epsilon_{max}(n)} - \sqrt{\epsilon_{min}(n)}}
\end{equation}
The value of $\alpha$ is $0.001$ in the $P^\prime_r(n)$ function, providing sufficient enhancement of the magnetization, while the constant $S(T)$ is used as a scale-matching parameter to compare these magnetization results with paramagnetic magnetizations (above the GP region) and to smooth the singularity often observed as a broad feature in experiments\cite{ref-16}. The value of is $S(T)$ set as
\begin{equation}
S(T)= \frac{0.001\beta}{7000 - m(T)}
\end{equation}
\\
Finally, for the calculation of the magnetization at a temperature higher than $T_g$, one may use the following expressions.
\begin{equation}
\resizebox{.97\hsize}{!}
{$M_p^p = n_p^{pt} \mu_B \tanh \frac{\mu_B (H + \lambda_w M_p^p)}{k_B T} \text{, } n_p^{pt} \propto  \sum_{n=1}^{n_{max}} n P_r (n)$}
\end{equation}
With this, we estimate the inverse susceptibility $\chi^{-1}$ at different temperatures for a magnetic field of $H = 0.5 $ Tesla and for different values of $p$, as shown in \textbf{Figure 2a}. The value of $T_g$ was fixed at $200~\text{kelvin}$ for all these calculations. As predicted theoretically and observed in experiments (\textbf{Figure 1b}), we find a sharp downward deviation from the linear Curie–Weiss behavior of the inverse susceptibility below $200~\text{kelvin}$. The strength of this downward deviation becomes more pronounced as the disorder strength $p$ increases. Following a procedure similar to that described in the introductory section, the strength of this deviation is estimated using an expression of the form $\chi^{-1} = (T - T_c)^{1- \lambda_g(T)}$ near the global FM transition temperature. The estimated values of $\lambda_{g}(T)$ for different disorder strengths are plotted in the inset of \textbf{Figure 2a}. As expected, the strength of the GP, indicated by the deviation of $\lambda_{g}(T)$ from 1 in the paramagnetic region, increases with increasing disorder strength. \textbf{Figure 2b} shows the effect of the magnetic field on the GP behavior for a fixed disorder strength of $p = 0.5$. A suppression of the susceptibility $\chi$, or, in other words, an increase of $\chi^{-1}$ with magnetic field, is observed. These calculated behaviors (\textbf{Figures 2a and 2b}) agree well with the experimentally observed GP behavior shown in \textbf{Figure 1b}. However, as pointed out in the introductory section, these behaviors alone do not guarantee the formation of a genuine GP. To further elucidate the nature of the GP, we calculate the variation of magnetization $(M)$ with magnetic field $(H)$ at several fixed temperatures for a fixed disorder strength of $p = 0.5$, as shown in \textbf{Figure 2c}. Notably, the magnetization does not follow the typical Brillouin or Langevin behavior expected for ferromagnetic clusters; instead, the variation in the GP region below $200~\text{kelvin}$ is best described by a power-law behavior of the form $M = c + x H^{\lambda_H}$. The estimated values of $\lambda_H$ from the fitting of the curves are shown in \textbf{Figure 2d}. As expected, the value of $\lambda_H$ is near 1 in the PM region and deviates downward $(\lambda_H < 1)$ as the temperature drops below the GP temperature. Interestingly, the deviation of $\lambda_H$ closely resembles the deviation from the CW behavior shown in \textbf{Figures 2a, 2b, and 1b}. Since the value of $\lambda_H$ is below 1 in the GP region, the differential susceptibility $\chi_{d} = dM/dH$, shown in the inset of \textbf{Figure 2c} for a particular temperature, diverges near $H = 0$. Therefore, the free energy and magnetization are indeed nonanalytic at $H=0$ within the GP region of a 3D Ising ferromagnetic system, as predicted originally\cite{ref-1}.\\
\subsection {Antiferromagnetic system}
Establishing the nonanalytic behavior of magnetization in an Ising FM sample, we proceed to examine the behavior of the uniform and staggered magnetization for a spin-1/2 3D Ising AFM system, which is not yet well understood in 3D Ising AFM systems\cite{ref-18}. Before analyzing the behavior of the AFM sample, it must be noted that the effect of dilution in an AFM system can be very different from that in an FM sample. However, noting that in many systems a similar suppression of the AFM transition temperature $T_n$ as that of the FM $T_c$ system has been observed, we may apply GP physics to the AFM system as well\cite{ref-35}. However, it also needs to be confirmed whether the GLW theory and rare region physics are appropriate for a disordered AFM system. In this regard, we note that in many samples, slightly below and near the percolation threshold, the formation of a cluster glass ground state has been observed\cite{ref-36,ref-37,ref-38}. This observation implies that dilution effects can indeed fragment (at least in certain systems) the AFM matrix into clusters, and that the GLW theory, along with rare region physics, may be applied to such cases \cite{ref-36,ref-37,ref-38}. To calculate the uniform magnetization $M_{un}^p$ in the paramagnetic region (above $T_g$) under a uniform magnetic field $H_{un}$ above the GP region, a similar expression to that in (\textbf{Equation 14}) was used for the two sublattices, A and B. The global AFM transition temperature $T_n$, required to estimate $\lambda_w$ in the presence of disorder $p$, is determined using the same approach as in the FM system. The Weiss field equation for the present scenario can be expressed as:
\begin{equation}
\begin{aligned}
&M_{Ap}^p = n_p^{p} 0.5 \mu_B \tanh \frac{\mu_B (H_{un} - \lambda_w M_{Bp}^p)}{k_B T}\\
&M_{Bp}^p = n_p^{p} 0.5 \mu_B \tanh \frac{\mu_B (H_{un} - \lambda_w M_{Ap}^p)}{k_B T}\\
& M_{un}^p = M_{Ap}^p + M_{Bp}^p \text{ and } n_p^p \propto  \sum_{n=1}^{n_{max}} n P_r (n)
\end{aligned}
\end{equation}
The magnetization in the GP region is estimated by approximately accounting for the reduction of magnetization due to AFM cluster formation, using an empirical expression of the form
\begin{equation}
M_{un}^{gp} = M_{un}^p - S (M_{un}^{pc} - M_{un}^{afm})
\end{equation}
Here, $M_{un}^p$ represents the total paramagnetic magnetization in the absence of $T_g$, given by an expression of the form (\textbf{Equation 15}), whereas the term in brackets quantifies the reduction due to cluster formation, including the effects of gaps discussed above. $S$ is a proportionality constant used to prevent negative values. The terms $M_{un}^{pc}$ and $M_{un}^{afm}$ can be expressed as\\
\begin{widetext}
\begin{equation}
M_{un}^{pc} (T,H) = \sum_{n=m(T)}^{n_{max}} P^\prime_r(n) \int_{\epsilon_{min}(n)}^{\epsilon_{max}(n)} P_n(\epsilon) M_{c-un}^p (n,H,T) \tanh \frac{M_{c-un}^p (n,H,T) H}{\epsilon}
\end{equation}
\end{widetext}
\begin{widetext}
\begin{equation}
M_{un}^{afm} (T,H) = \sum_{n=m(T)}^{n_{max}} P^\prime_r(n) \int_{\epsilon_{min}(n)}^{\epsilon_{max}(n)} P_n(\epsilon) M_{c-un}^{afm} (n,H,T) \tanh \frac{M_{c-un}^{afm} (n,H,T) H}{\epsilon}
\end{equation}
\end{widetext}
The magnetization can be obtained by solving the Weiss equations using the Newton–Raphson method\cite{ref-39}. The Weiss equations are
\begin{equation}
\begin{aligned}
&M_{Ap} = 0.5 n \mu_B \tanh \frac{\mu_B (H_{un} - \lambda_w M_{Bp})}{k_B T}\\
&M_{Bp} = 0.5 n \mu_B \tanh \frac{\mu_B (H_{un} - \lambda_w M_{Ap})}{k_B T}\\
& M_{c-un}^p (n,H,T) \text{ or } M_{c-un}^{afm} (n,H,T) = M_{Ap} + M_{Bp}
\end{aligned}
\end{equation}
\begin{figure}[t]
\begin{center}
\includegraphics[width=1.0\columnwidth]{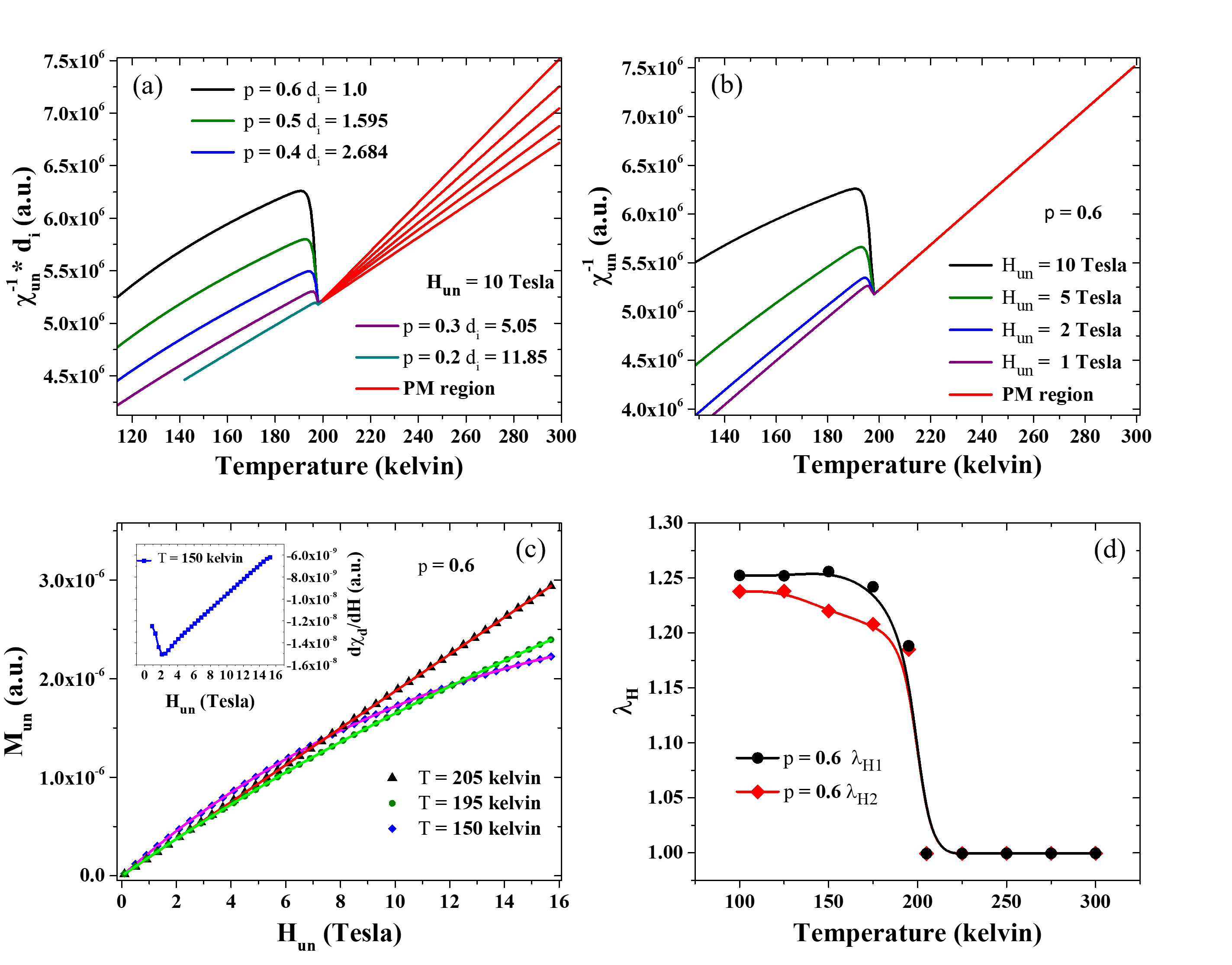}
\caption{(a) The calculated inverse susceptibility, $\chi_{un}^{-1}$, of an AFM system, both above and below $T_g$, is plotted as a function of temperature for different disorder strengths, $p$. A magnetic field of $H_{un} = 10$ Tesla was used for these calculations. In order to compare the $\chi_{un}^{-1}$ values calculated at different disorder strength, $p$, the calculated $\chi_{un}^{-1}$ values were multiplied by a constant factor $d_i$ for better comparison. Also note that, restricted by the maximum cluster size, the calculations were performed up to $\sim 198~\text{kelvin}$; the region between $\sim 198~\text{kelvin}$ and $200~\text{kelvin}$ is an extrapolation. (b) shows the variation of $\chi_{un}^{-1}$ at different fields for a fixed disorder strength of $p = 0.6$ (c) The calculated variation of $M_{un}$ with $H_{un}$ at a few selected temperatures for a fixed disorder strength of $p = 0.6$ is shown. The solid lines represent fits using a power-law behavior. (d) The estimated values of the exponent, obtained from the fitting of the power-law variation of $M_{un}$ versus $H_{un}$, are shown for a disorder strength of 0.6.}
\label{fig3}
\end{center}
\end{figure}
\begin{figure}[t]
\begin{center}
\includegraphics[width=1.0\columnwidth]{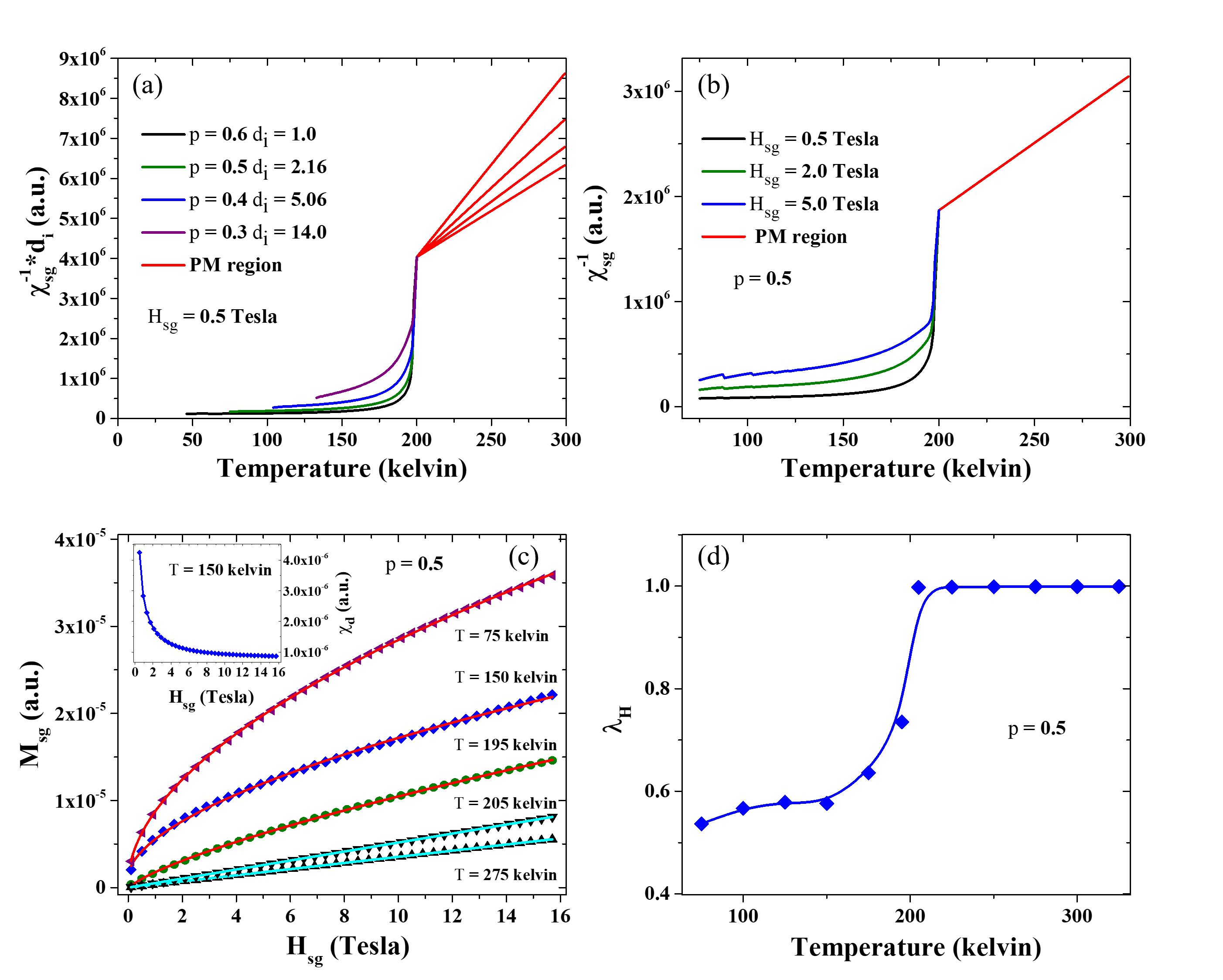}
\caption{(a) The calculated inverse susceptibility, $\chi_{sg}^{-1}$, of an AFM system, both above and below $T_g$, is plotted as a function of temperature for different disorder strengths, $p$. A magnetic field of $H_{sg} = 0.5$ Tesla was used for these calculations. In order to compare the $\chi_{sg}^{-1}$ values calculated at different disorder strength, $p$, the calculated $\chi_{sg}^{-1}$ values were multiplied by a constant factor $d_i$ for better comparison. Also note that, restricted by the maximum cluster size, the calculations were performed up to $\sim 198~\text{kelvin}$; the region between $\sim 198~\text{kelvin}$ and $200~\text{kelvin}$ is an extrapolation. (b) shows the variation of $\chi_{sg}^{-1}$ at different fields for a fixed disorder strength of $p = 0.5$ (c) The calculated variation of $M_{sg}$ with $H_{sg}$ at a few selected temperatures for a fixed disorder strength of $p = 0.5$ is shown. The solid lines represent fits using a power-law behavior. (d) The estimated values of the exponent, obtained from the fitting of the power-law variation of $M_{sg}$ versus $H_{sg}$, are shown for a disorder strength of 0.5.}
\label{fig4}
\end{center}
\end{figure}
Note that the difference in the values of $M_{c-un}^p (n,H,T)$ and $M_{c-un}^{afm} (n,H,T)$ arises from $\lambda_w$ and the temperature. At a fixed $T$, the first expression corresponds to clusters in the paramagnetic region above the transition temperature $T_n$ ($\lambda_w = \lambda_w$), whereas the second corresponds to clusters that are AFM and below $T_n^r$ ($\lambda_w = \lambda_{wc}$), as estimated using the finite-size scaling method discussed previously. Note that $ M_{c-un}^p (n,H,T)$ is a completely empirical term, suggesting that the clusters follow the quantum GP gap distribution theory even when they are paramagnetic. However, the use of this hypothesis provides an empirical way to estimate the reduction of magnetization upon cluster formation. The proportionality constant and the scale-matching parameter are set as $S=10^{-6}\beta$ to ensure non-negative values of magnetization within the simulation range. The remaining parameters are the same as those used in the FM system.\\
We estimate the inverse susceptibility $\chi_{un}^{-1}$ at different temperatures for a magnetic field of $H_{un} = 10$ Tesla and for different values of $p$, as shown in \textbf{Figure 3a}. The value of $T_g$ was fixed at $200~\text{kelvin}$ for all these calculations. Note that $\chi_{un}^{-1}$ reflects contributions both from clusters that are not ordered at a particular temperature and from clusters that are ordered antiferromagnetically. We find that the signature of deviation from linear CW behavior is extremely weak for low disorder and low field values (\textbf{Figures 3a and 3b}). However, with increasing disorder and $H_{un}$, a sharp deviation from linear behavior becomes prominent. However, we find that, unlike in the FM system, the deviation is upward, suggesting a suppression of susceptibility upon cluster formation, as expected for AFM clusters. With increasing $H_{un}$, further suppression of susceptibility is observed, which is similar to the behavior in the FM system. \textbf{Figure 3c} shows the variation of the uniform magnetization $M_{un}$ with different $H_{un}$, calculated at temperatures both below and above $T_g$. The behavior of the magnetization above $T_g$ is paramagnetic-like. However, below $T_g$, the behavior becomes nonlinear, as expected in the presence of clusters. This nonlinear behavior is best described using an expression of the following form $M_{un} = c + (x H_{un}^{\lambda_{H2}} - y H_{un}^{\lambda_{H1}})$. The values of the exponents are shown in \textbf{Figure 3d}. Note that, unlike in the FM system, the values of the exponents are non-integer and greater than 1, and hence the differential susceptibility $\chi_d = dM_{un}/dH_{un}$ does not diverge at $H_{un} = 0$. However, given the fractional nature of the exponents, the higher-order derivatives of the magnetization $d\chi_{d}/dH_{un}$ do diverge, as shown in the inset of \textbf{Figure 3c}. These results suggest that the uniform magnetization is a nonanalytic function of $H_{un}$ at $H_{un} = 0$ for a 3D Ising AFM system in the GP region. To the best of our knowledge, these unusual behaviors—including the upturn deviation of $\chi_{un}^{-1}$ and the power-law behavior of $M_{un}$ —have been observed in only a very few rare samples\cite{ref-40,ref-41,ref-42,ref-43}, experimentally, partly because, unlike in FM systems, the deviation from linear Curie–Weiss behavior in AFM systems is very weak at low fields and low disorder. However, even in those samples reporting an upward deviation, $(\theta_{CW})$ is found to be positive\cite{ref-40,ref-41,ref-42,ref-43} and therefore may not signify an AFM GP. Consequently, we suggest that the existence of an AFM GP still needs to be established experimentally. However, it must be noted that, in a diluted AFM, magnetic interactions can be further complicated by the presence of uncompensated surface spins and boundary effects \cite{ref-44,ref-45}, effectively forming a cluster in which the inner core is AFM, whereas the outer shell is either superparamagnetic or FM-like. Such a description requires further investigation and is beyond the scope of the present discussion. On the other hand, it should be noted that the uniform magnetization, which is mostly measured in magnetization experiments, does not represent the order parameter for an AFM system and therefore cannot be directly related to the free energy. The staggered magnetization is the true order parameter for an AFM system. Furthermore, as will be discussed subsequently, unlike the uniform magnetization—which requires an empirical expression to describe the behavior in the GP region—the GP behavior in an AFM system can be better understood in terms of staggered magnetization, providing a cleaner and more straightforward description without involving empirical expressions.\\

We calculate the behavior of the staggered magnetization ($M_{sg}$) in a randomly disordered spin-1/2 3D Ising AFM system. For this purpose, we use exactly the same method described for a FM system through \textbf{Equations 1 to 14}, except with the following changes. First, all calculations are performed using a staggered magnetic field ($H_{sg}$). Second, the Weiss equations are appropriately modified to include the staggered field and the AFM sublattices A and B.
\begin{equation}
\begin{aligned}
&M_{Ap} = 0.5 n \mu_B \tanh \frac{\mu_B (H_{sg} - \lambda_w M_{Bp})}{k_B T}\\
&M_{Bp} = 0.5 n \mu_B \tanh \frac{\mu_B (-H_{sg} - \lambda_w M_{Ap})}{k_B T}\\
& M_{sg} = M_{Ap} - M_{Bp}
\end{aligned}
\end{equation}
The value of $\lambda_w$ should be chosen accordingly for paramagnetic spins ($\lambda_w$) or clusters ($\lambda_{wc}$), depending on the strength of their interactions. The rest of the parameters and methods are exactly the same as those of a FM system. \textbf{Figure 4 (a–d)} shows the behavior of the staggered inverse susceptibility $\chi_{sg}^{-1}$, staggered magnetization $M_{sg}$, etc. Note that the behavior is approximately similar to that of a disordered FM system exhibiting a GP region, and the results suggest that the free energy and magnetization are nonanalytic at $H_{sg} = 0$ within the GP region of a 3D Ising AFM system, similar to a FM system.\\

\subsection {Ferrimagnetic system}
\begin{figure}[t]
\begin{center}
\includegraphics[width=1.0\columnwidth]{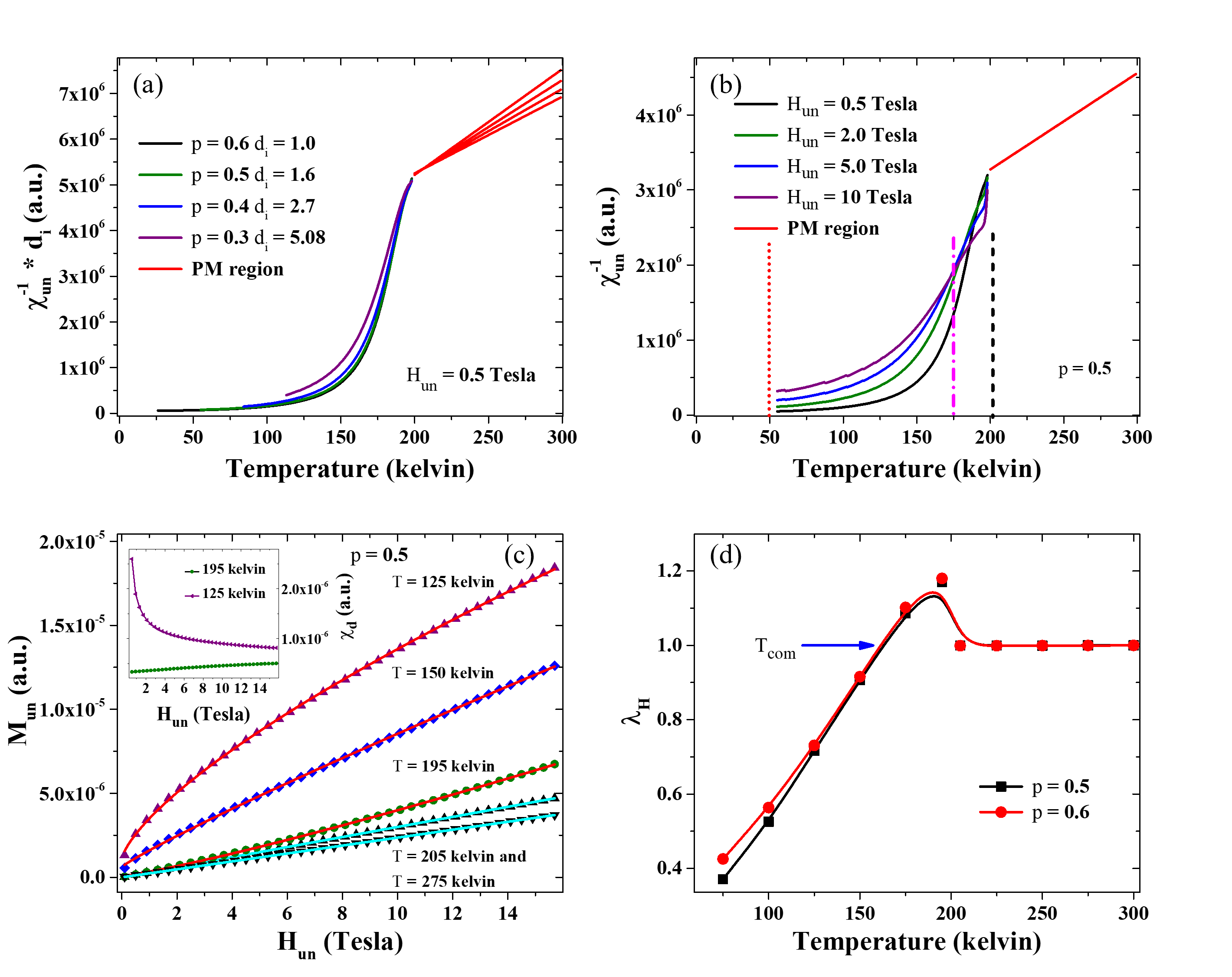}
\caption{(a) The calculated inverse susceptibility, $\chi_{un}^{-1}$, of a FiM system, both above and below $T_g$, is plotted as a function of temperature for different disorder strengths, $p$. A magnetic field of $H_{un} = 0.5$ Tesla was used for these calculations. In order to compare the $\chi_{un}^{-1}$ values calculated at different disorder strength, $p$, the calculated $\chi_{un}^{-1}$ values were multiplied by a constant factor $d_i$ for better comparison. (b) shows the variation of $\chi_{un}^{-1}$ at different fields for a fixed disorder strength of $p = 0.5$ (c) The calculated variation of $M_{un}$ with $H_{un}$ at a few selected temperatures for a fixed disorder strength of $p = 0.5$ is shown. The solid lines represent fits using a power-law behavior. (d) The estimated values of the exponent, obtained from the fitting of the power-law variation of $M_{un}$ versus $H_{un}$, are shown for two different disorder strengths.}
\label{fig5}
\end{center}
\end{figure}
Finally, we extend this method to a 3D spin-$1/2$ Ising FiM system. A FiM system can be modeled from an AFM system by removing some of the spins from the B sublattice while keeping all the spins in the A sublattice intact. For this calculation, we assume that the A sublattice has all its spins intact, while 50$\%$ of the spins in the B sublattice are removed. Therefore, if a cluster has 100 spins, approximately 67 are up spins and approximately 33 are down spins. Interestingly, unlike a pure AFM sample, a FiM sample exhibits spontaneous magnetization; therefore, the uniform magnetization acts as the order parameter for a FiM sample. The method for calculating the magnetization is the same as that described earlier for a FM sample. The only change made is in appropriately formulating the Weiss field equation for a FiM sample.
\begin{equation}
\begin{aligned}
&M_{Ap} = 0.667 n \mu_B \tanh \frac{\mu_B (H_{un} - \lambda_{w1} M_{Bp})}{k_B T}\\
&M_{Bp} = 0.333 n \mu_B \tanh \frac{\mu_B (H_{un} - \lambda_{w2} M_{Ap})}{k_B T}\\
& M_{un} = M_{Ap} + M_{Bp}
\end{aligned}
\end{equation}
The value of $\lambda_w$ should be chosen accordingly for paramagnetic spins ($\lambda_w$) or clusters ($\lambda_{wc}$), depending on the strength of their interactions. Additionally, the scale-matching parameter $S(T)$, used to compare these magnetization results with paramagnetic magnetization (above the GP region) as well as to smooth the singularity often observed as broad behavior in experiments, is defined as
\begin{equation}
S(T)= \frac{0.01\beta}{7000 - m(T)}
\end{equation}
Note that the removal of some spins from the B sublattice may introduce an additional exponent factor in $P_r(n)$ and $P^\prime_r(n)$; however, the overall behavior of the magnetization is not expected to change, as the exponential distribution remains intact. Therefore, we proceed with the usual method described in the earlier sections.\\
We estimate the inverse uniform susceptibility ($\chi_{un}^{-1}$) at different temperatures under a uniform magnetic field of $H_{un} = 0.5 $ Tesla for different values of $p$, as shown in \textbf{Figure 5a}. The value of $T_g$ is fixed at $200~\text{kelvin}$ for all these calculations. We observe a sharp downturn deviation from the linear CW behavior of $\chi_{un}^{-1}$ below $200~\text{kelvin}$, the magnitude of which increases slightly with increasing disorder strength $p$. This behavior is similar to that of a FM system. \textbf{Figure 5b} shows the behavior of $\chi_{un}^{-1}$ as a function of temperature at different magnetic fields $H_{un}$ for a fixed disorder value $p = 0.5$. Interestingly, unlike FM or AFM systems, we observe an unusual behavior of $\chi_{un}^{-1}$ at different values of $H_{un}$. Specifically, the region close to $T_g$ (marked by black and magenta lines) shows an enhancement of the susceptibility with increasing $H_{un}$. As we move deeper into the GP region, this behavior changes, and the usual suppression of susceptibility with increasing $H_{un}$ is observed, similar to that in FM or AFM systems. To understand this unusual behavior, we plot the uniform magnetization $M_{un}$ as a function of $H_{un}$ at a few selected temperatures for a fixed disorder strength $p = 0.5$, as shown in \textbf{Figure 5c}. For temperatures above $T_g$, the usual paramagnetic behavior is observed. However, just below $T_g$, the system shows an unusual dependence of $M_{un}$ on $H_{un}$, which follows a power-law form, $M_{un} = c + xH_{un}^{\lambda_H}$ , with an exponent greater than one. As the temperature is further lowered, the exponent becomes less than one, similar to that of a FM system. The behavior of $\lambda_H$ at different temperatures is shown in \textbf{Figure 5d} for two disorder strengths, $p = 0.5$ and $p = 0.6$. This behavior has a pronounced effect on the differential susceptibility $\chi_d = dM_{un}/dH_{un}$. Since $\chi_d \propto H_{un}^{\lambda_H - 1}$, a value of $\lambda_H$ greater than one induces an increase of $\chi_d$ with increasing $H_{un}$. On the other hand, if $\lambda_H$ is less than one, $\chi_d$ decreases with increasing $H_{un}$. These opposite behaviors of $\chi_d$ (shown in the inset of \textbf{Figure 5c}) result in an enhanced or suppressed region in the behavior of $\chi_{un}^{-1}$, as shown in \textbf{Figure 5b}. Further, as stated before for AFM and FM systems, a fractional but greater-than-one $\lambda_H$ leads to a divergence of the higher derivatives of $M_{un}$, whereas a fractional $\lambda_H$ less than one leads to a divergence of $\chi_d$ itself at $H_{un}=0$, as shown in the inset of \textbf{Figure 5c}. Therefore, we may conclude that, similar to a FM system, the free energy and $M_{un}$ of a FiM system are nonanalytic functions of $H_{un}$ at $H_{un}=0$ for most temperatures below $T_g$. However, interestingly, this crossover of $\lambda_H$ from greater than one to less than one leads to a temperature ($T_{com}$) below $T_g$ where $\lambda_H$ is precisely one. At this compensation point $T_{com}$, marked by a blue arrow in \textbf{Figure 5d}, the free energy and $M_{un}$ are analytic functions of $H_{un}$, even when the temperature is below $T_g$.

\section{Conclusion}

In conclusion, we have attempted to understand the behavior of magnetization within the GP region in FM, AFM, and FiM systems. While the suggested methods are slightly empirical in selected cases due to the inherent difficulties associated with the subject, they do provide a framework for identifying GP behavior in different systems. We hope that, alongside experimentally studied FM systems, the presented method will also help in identifying AFM and FiM-type GP behavior experimentally.

\section{Acknoledgement}

The author thanks the Indian Institute of Science, Bangalore, for providing the facilities to conduct this work. The author thanks Soumya Kanti Ganguly for useful discussions. Parts of the manuscript was proofread for grammatical accuracy using ChatGPT, an AI-based language tool.

\section{Data availability}

The data that support the findings of this study are available from the corresponding author upon reasonable request.

%\newpage

%\section{Supporting Information}

\end{document}